\documentclass{elsart5p}

\usepackage{natbib}

\usepackage{graphicx}

\usepackage{amssymb}

\journal{New Astronomy}

\begin{document}

\begin{frontmatter}



\title{Probing the feeding and feedback of AGN through 
molecular line maps}


\author{S. Garc\'{\i}a-Burillo}
\ead{s.gburillo@oan.es}
\address{Observatorio Astron\'omico Nacional-OAN, Observatorio de Madrid, Alfonso XII, 3, E-28014, Madrid, SPAIN}
\author{F.~Combes}
\ead{Francoise.combes@obspm.fr}
\address{Observatoire de Paris, LERMA, 61, Av. de l'Observatoire, 75014 Paris, FRANCE}
\author{A.~Usero, J.~Graci\'a-Carpio}
\ead{a.usero@oan.es, j.gracia@oan.es}
\address{Observatorio Astron\'omico Nacional-OAN, Observatorio de Madrid, Alfonso XII, 3, E-28014, Madrid, SPAIN}

\begin{abstract}

Current mm-interferometers can provide a complete view
of the distribution and kinematics of molecular gas in the circumnuclear disks 
of nearby galaxies. High-resolution CO maps are paramount in order to track 
down the feeding of active nuclei and quantitatively address the issue of how 
and for how long nuclear activity can be sustained in galaxies. Going beyond 
CO mapping, the use of more specific molecular tracers of dense gas can probe 
the feedback influence of activity on the chemistry and energy balance/redistribution 
in the interstellar medium of nearby galaxies, a prerequisite to interpret how 
feedback may operate at higher redshift galaxies. In this context we present the 
latest results issued from the NUclei of GAlaxies (NUGA) project, a high-resolution 
(0.5$^{\prime\prime}$-1$^{\prime\prime}$) CO survey of low luminosity AGNs conducted with the IRAM interferometer. 
The efficiency of gravity torques as a mechanism to account for the feeding of low luminosity 
AGNs (LLAGNs) can be analyzed. We discuss an evolutionary scenario in which gravity 
torques and viscosity act in concert to produce recurrent episodes of activity during 
the typical lifetime of any galaxy. We also present the results of an ongoing survey 
allying the IRAM 30m telescope with the Plateau de Bure Interferometer (PdBI), devoted 
to probe the feedback of activity through the study of the excitation and chemistry of 
the dense molecular gas in a sample of nearby AGNs and ULIRGs as well as in a prototypical 
high-redshift QSO.

\end{abstract}


\end{frontmatter}

\section{Probing AGN feeding}

Nuclear activity in galaxies is understood as the result of the feeding of supermassive black holes. Active Galactic Nuclei (AGN) must be fuelled with material  which lies in the disk of the host galaxy originally far away from the gravitational influence of the black hole. This implies that gas must loose virtually all of its angular momentum to come from $\sim$kpc-scale distances to the inner $\sim$pc on its way to the nucleus . While for high luminosity AGNs (HLAGNs), there is a well established correlation between the presence of $\sim$kpc-scale non-axisymmetric perturbations (large-scale bars, and interactions) and the onset of activity, the case for a similar correlation in low luminosity AGNs (LLAGNs) is weak \citep{mol95, kna00, sch01, com03}. The quest of a {\it universal} mechanism for AGN feeding in LLAGNs is probably complicated by the fact that the AGN duty cycle ($\sim$10$^{7-8}$ years) might be shorter than the lifetime of the feeding mechanism itself. Furthermore, it is unclear whether a single mechanism or, alternatively, a hierarchy of mechanisms combining in due time can explain the onset of activity in LLAGNs. In summary, in spite of all the theoretical and observational efforts, the LLAGN fueling problem has thus far remained an unsolved problem \citep[e.g., see reviews by][]{com03, mar04}.

\subsection{The NUGA project}

In order to probe the critical scales of AGN feeding ($<$100~pc), i.e., the scales on which secondary modes embedded in kpc-scale perturbations are expected to take over, we embarked on a high-resolution ($\sim$0.5''-1'') CO survey of a sample of 25 LLAGNs carried out with the Plateau de Bure Interferometer (PdBI) (fully described by \citet{gb03a, gb03b}). The CO line maps provided by the NUclei of GAlaxies (NUGA) project in 12 LLAGNs (including Transition objects, LINERs and Seyferts of types 1 and 2) can probe the distribution and kinematics of molecular gas in the circumnuclear disks (CND) of these galaxies with unprecedented resolution and sensitivity compared to previous interferometer surveys of LLAGNs \citep{jog01, koh01}. 

\begin{figure*}[!bth]
  \centering
    \includegraphics[width=12cm]{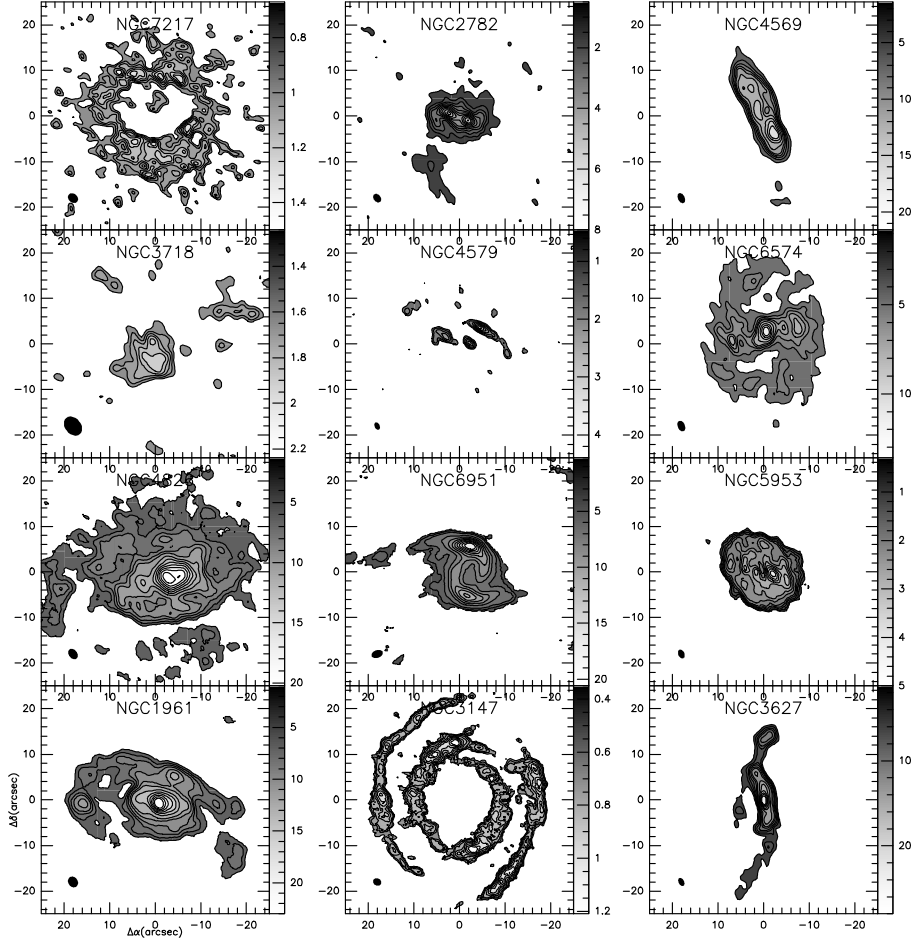}
     \caption{$^{12}$CO(1--0) maps obtained for the 12 galaxies belonging to the core sample of NUGA. A wide range of morphologies are identified in the circumnuclear disks of these galaxies. This suggests that several mechanisms may be at work to feed the central engines of LLAGNs \citep{gb05}. Integrated intensities in Jy~km~s$^{-1}$ are indicated by grey scale levels particularized for each galaxy.}
         \label{nuga-1}
   \end{figure*}

Figure~\ref{nuga-1} shows the CO(1--0) maps obtained in the 12 galaxies of the core sample of NUGA. This subsample includes those
galaxies for which we reach the highest sensitivities and spatial resolutions in the CO 1--0 and 2--1 maps \citep{gb03b, com04, kri05}. The large diversity of morphologies identified in the CND of these galaxies (including two-arm spirals, m=1 instabilities, rings, gas bars and axisymmetric or mostly featureless disks) is already suggestive of an evolutionary scenario where several mechanisms (and not just one single universal mechanism) can cooperate in due time to feed the central engines of LLAGNs. 

A detailed case-by-case study of the distribution and the kinematics of molecular gas of all the galaxies in our sample on the scales which are critical for AGN feeding can be interpreted in terms of evidence of ongoing feeding. This case-by-case in-depth study is the approach adopted to take full advantage of the high quality of the NUGA maps. In particular, we have also information on the stellar potentials of the majority of NUGA galaxies. This information, obtained through available HST and ground-based optical/NIR images of the sample, is used to determine the gravitational torques exerted by the derived stellar potentials on the gaseous disk. The efficiency with which gravitational torques  drain the angular momentum of the gas depends first on the strength of the non-axisymmetric perturbations of the potential (m$>$0) but, also, on the existence of significant phase shifts between the gas and the stellar distributions. Therefore, the estimate of these phase shifts necessarily requires the availability of images of comparably high spatial resolution ($\leq$0.5$^{\prime\prime}$ in the case of NUGA) showing the distribution of the stars and the gas. 

\subsection{Tracking down gravity torques}

\citet{gb05} have focused on the study of gravitational torques in a subset of NUGA galaxies, which span the range of the different activity classes within our sample: NGC\,4321 (transition object: HII/LINER), NGC\,4826 (transition object: HII/LINER),
NGC\,4579 (LINER 1.9/Seyfert 1.9) and NGC\,6951(Seyfert 2). Our calculations allow us to derive the characteristic time-scales for gas flows and discuss whether torques from the stellar potentials are efficient enough to drain the gas angular momentum in the inner 1~kpc of these galaxies. After estimating the role of stellar gravitational torques, we have investigated whether other mechanisms are required to explain the low level of nuclear activity in the galaxies analyzed thus far \citep{gb05}. 

As fully explained in \citet{gb05}, gravitational forces were derived in the plane of the galaxy, using NIR images to calculate a fair estimate of the gravitational potential (NIR maps are here implicitly assumed as mostly extinction free). We also assumed that the mass budget is dominated by the stellar contribution and thus purposely ignore gas self-gravity (however, the role of gas self-gravity certainly deserves further scrutiny in some cases). Furthermore, we hypothesized that the M/L ratio is roughly constant, and determined its best value by fitting the rotation curve constrained by the CO observations. From the 2D force field ($F_x$,$F_y$) we derived the torques per unit mass at each location ($t(x,y)$=$x~F_y-y~F_x$). This torque field, by definition, is independent of the present gas distribution in the plane.

 The crucial step in this method consists of using the torque field to derive the angular momentum variations and the associated flow time-scales.  To do that we first assume that the measured gas column density ($N(x,y)$) derived from a CO intensity map at each offset in the galaxy plane is a good estimate of the probability of finding gas at this location at the present time. In this statistical approach, we implicitly average over all possible orbits of gaseous particles and take into account the time spent by the gas clouds along the orbits. 
The torque field is then weighted by $N(x,y)$ at each $(x,y)$-position to derive the time derivative of the local angular momentum surface density $dL_s(x,y)/dt$=$N(x,y) \times t(x,y)$. To quantitatively derive the gas flows induced by these angular momentum variations we produce azimuthal averages of $dL_s(x,y)/dt$ at each radius using $N(x,y)$ as the actual weighting function. These averages represent the global variation of the specific gas angular momentum occurring at this radius ($dL/dt~\vert_\theta$). Finally, the time-scales for gas inflow/outflow can be derived by estimating the average fraction of angular momentum transferred in one rotation ($dL/L$), where $L$ is estimated by the 
axisymmetric value $L_\theta=R\times v_{rot}$.

  \begin{figure*}[!bth]
  \centering
    \includegraphics[width=12cm]{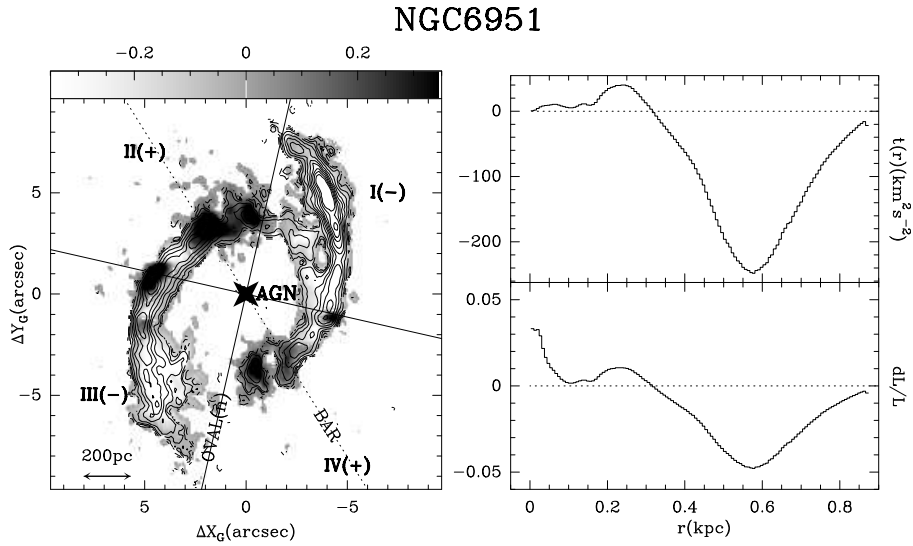}
     \caption{The gravity torque budget in NGC~6951 derived from the CO NUGA map and the J-band HST image of the galaxy \citep{gb05}. The torque map (left panel) shows the expected butterfly pattern of positive/negative torques driven by the nuclear oval (OVAL(n)). The derived radial profile of the angular momentum derivative (t(r)) and the fraction of angular momentum transferred in one rotation (dL/L), both show that gravity torques can drain angular momentum from the gas down to the ring. In contrast, stellar torques become positive inside the ring (right panels).}
         \label{nuga-2}
   \end{figure*}

The gravity torque method can be applied to derive gas flow time-scales in galaxies from the outer disk (using HI as a tracer of gas at large radii) down to the nucleus (where CO lines are the best choice to trace the neutral gas phase). The extension of this analysis to the whole NUGA sample is currently underway.

\subsection{A new scenario for self-regulated activity in LLAGNs}

The results obtained from the analysis of stellar torques have revealed a puzzling feeding budget in the CND of
the subset of NUGA targets studied by \citet{gb05}. The picture emerging in the two Seyferts/LINERs (NGC~6951
and NGC~4579) indicates that AGN feeding may proceed in two steps. In a first step, gravity torques
help driving the gas inwards and feeding a nuclear starburst on scales of $\sim$a few 100~pc. On smaller scales, however, stellar torques play no role in AGN fueling in the current epoch: torques on the gas are not negative all the way to the center, but on
the contrary they become positive and quench the feeding (see Figure~\ref{nuga-2} adapted from \citet{gb05}). This may explain why molecular gas seems to 'avoid' the inner 200-300~pc of NGC~6951 and NGC~4579 where we measured M$_{gas}<$a few$\times$10$^{5}$--10$^{6}$M$_{\odot}$. 

One possible explanation of this puzzle is that the responsible agent in the stellar potential 
could be as short lived as an individual AGN episode (hence: $\leq$10$^{7-8}$ years). The feeding phase
from 100~pc to 1~pc could be so short that the smoking gun evidence in the
potential is missing. Alternatively, this {\it temporary inability} of gravitational torques
could be overcome by other mechanisms that, over time, become competitive with non-axisymmetric
perturbations. Among the different mechanisms usually cited in the literature, dynamical friction
and viscous torques have been invoked to help AGN feeding on $\sim$100~pc scales.

Dynamical friction of Giant Molecular Clouds (GMCs) on the stellar bulge is likely very inefficient due to the strong clumpiness of molecular gas on a wide range of spatial scales: therefore, the ensuing gas flow time-scales, estimated by the classical Chandrasekhar's formula \citep{cha43} (which implicitly assumes a highly concentrated point-like source distribution for the GMC) are severely underestimated. On the contrary, {\it moderate} viscosity can counterbalance the positive gravity torques of a {\it weak} bar inside the ILR if it acts on contrasted nuclear rings. Based on the classical diffusion equation of \citet{pri81},  \citet{gb05} have estimated on a case-by-case basis the time-scale of viscous transport in the four LLAGNs analyzed above.  Indeed, the efficiency of viscous transport can be enhanced if the initial distribution of the disk is characterized by strong density gradients. This is precisely the case of nuclear contrasted rings, especially when these are located in the inner regions of galaxies (r$\sim$100--500~pc) like the ones observed in NGC~6951 (see Figure~\ref{nuga-2}) and NGC~4579. Furthermore, the galactic shear inferred from the steep $\Omega$-curve in these galaxies can still be very high on these scales; this also favors viscous transport.

 While bars can build up gas reservoirs towards the central regions of galaxies in the shape of nuclear rings, the dynamical
feed-back associated with these gas flows (through angular momentum exchange with the bar) can destroy or weaken the bars
\citep{bou02, bou05, com05}. When this happens, gravity torques are negligible and can make way for other competing mechanisms of gas transport, such as viscous torques, in an almost axisymmetric system. \citet{gb05} propose an evolutionary scenario in which the onset of nuclear activity can be understood as a recurrent phase during the typical lifetime of any galaxy. In this scenario the recurrence of activity in galaxies is indirectly related to that of the bar instabilities although the active phases are not necessarily coincident with the maximum strength of a single bar episode. These activity episodes are not expected to be strongly correlated with the phases of maximum strength for the bar, but they may appear at different evolutionary stages of the bar potential, depending on the balance between gravity torques and viscosity. The wide variety of morphologies revealed by the CO maps of the circumnuclear disks of NUGA targets corroborates that there is  no universal pattern associated with LLAGNs.

\section{Probing AGN feedback}

The huge energies injected into the circumnuclear gas reservoirs of AGNs through strong radiation fields (UV and X-rays) and mass flows (winds and jets) can determine to a large extent the physical and chemical status of the ISM in these galaxies.  In particular, molecular gas can be exposed to strong X-ray irradiation close to the central engine of active galaxies. Compared to  UV-photons (which are easily attenuated by dust opacity), hard X-rays (mostly absorbed by the gas itself) can extend their influence deeper into the clouds by their ability to penetrate huge gas column densities out to A$_v$=100-1000 \citep{mal96}. X-ray dominated regions (XDR) are revealed by an increase of the gas phase abundances of a certain set of ions, radicals and molecular species (including HCN, CN and NO)  \citep{lep96, mal96}. Going beyond pure gas-phase chemistry schemes, it has also been argued that X-rays could evaporate small ($\sim$10~\AA) silicate grains, increasing the fraction in gas phase of some refractory elements and subsequently enhancing the abundance of some molecules (e.g., SiO) in X-ray irradiated molecular gas \citep{voi91}.

The large HCN/CO intensity and abundance ratios measured in the nucleus of the Seyfert 2 galaxy NGC\,1068 by \citet{tac94} and \citet{ste94} constitute the first observational evidence that molecular gas chemistry can be shaped by activity.
\citet{use04} have detected strong SiO emission coming from the CND of NGC\,1068. The derived SiO abundances were seen to be significantly enhanced out to $\sim$10$^{-9}$. \citet{use04} made complementary observations of the CND of NGC\,1068 using the 30m telescope for eight molecular species, purposely chosen to explore the predictions of XDR models for molecular gas. These observations included several lines of CN, HCO, H$^{13}$CO$^{+}$, H$^{12}$CO$^{+}$, HOC$^{+}$, HCN, CS and CO. The first global analysis of the combined survey suggests that the bulk of the molecular gas emission in the CND of NGC\,1068 can be best interpreted as coming from a giant XDR created by the central engine. Of particular note, the results obtained from new interferometer observations of the nucleus of NGC~1068 have confirmed that the HCN molecule is significantly overabundant with respect to CO, HCO$^+$, HNC and CN (Garc\'{\i}a-Burillo et al., 2007, in prep.).  

 \begin{figure}[!th]
  \centering
    \includegraphics[width=8cm]{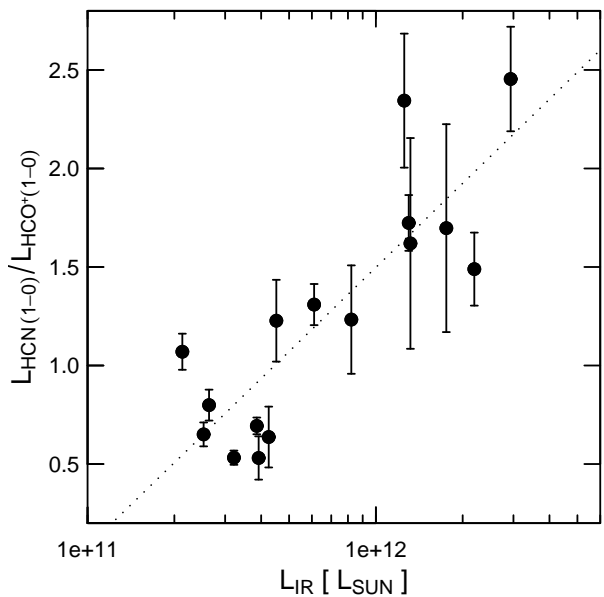}
     \caption{A puzzling trend between the HCN(1--0)/HCO$^{+}$(1--0) luminosity ratio and L$_{IR}$ has been identified by \citet{gra06} in a sample of LIRGs and ULIRGs. This casts doubts on the reliability of surveys made in just one molecular line (HCN or HCO$^+$) to quantitatively probe the dense molecular gas fraction of luminous infrared galaxies.}
         \label{ulirgs-1}
   \end{figure}

There is mounting evidence of overluminuous HCN lines in the nuclei of other nearby Seyferts \citep[e.g.,][]{koh01, koh05}, as well as in some luminous and ultraluminous infrared galaxies (LIRGs and ULIRGs) \citep{gao04, gra06, ima06}. The fact that these overluminous HCN lines can be plausibly explained by the enhancement of HCN abundances in the XDRs around AGNs (as seems to be the case of NGC~1068), casts founded doubts on 
the reliability of HCN as a {\it true} tracer of dense molecular gas in these galaxies. The possible caveats on the use of {\it only HCN observations} call for the use of alternative tracers of dense gas in AGNs, especially in those galaxies where AGNs are expected to be highly embedded as is the case of LIRGs, ULIRGs and, most particularly, of high redshift galaxies.

\subsection{Probing AGN feedback in ULIRGs}

The question of what is the underlying power source of LIRGs and ULIRGs is still a mostly debated and controversial issue. While extreme starbursts have been postulated as the dominant power source of the huge infrared luminosities ($L_{\rm{IR}}$) in many ULIRGs \citep{gen98}, a significant contribution from an embedded AGN cannot be excluded in some cases \citep{ris06, ima06}. 
LIRGs and ULIRGs possess large amounts of molecular gas as derived from CO(1--0) observations. \citet{san88} reported that the infrared--to--CO luminosity ratio in ULIRGs is anomalously high compared to that of normal galaxies and interpreted this result as evidence of the AGN power source scenario for ULIRGs. On the other hand, \citet{gao04} used HCN(1--0) observations to probe the dense molecular gas content of a sample of 65 nearby galaxies, including 25 LIRGs and 6 ULIRGs. Their results, showing a tight linear correlation between the IR and HCN luminosities over 3 orders of magnitude in $L_{\rm{IR}}$, were interpreted in terms of star formation as being the main power source in ULIRGs.
In Gao \& Solomon 's scenario, the {\it CO puzzle} for ULIRGs is apparently solved in a very simple and elegant way: ULIRGs just happen to have a higher fraction of dense molecular gas compared to normal galaxies and the need of an AGN source for their huge $L_{\rm{IR}}$ apparently vanishes. In terms of the star formation efficiency  (SFE), measured with respect to the fraction of dense gas derived from HCN(1--0) lines, normal galaxies and ULIRGs are alike.

 \begin{figure}[!th]
  \centering
    \includegraphics[width=8cm]{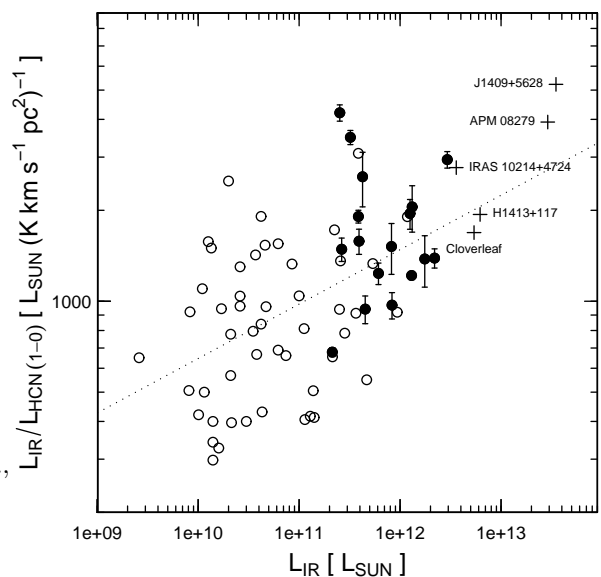}
     \caption{New HCN(1--0) observations (filled circles) obtained by Graci\'a-Carpio et al. 2007 (in prep) show that, contrary to the findings of \citet{gao04}, the star formation efficiency (SFE) measured with respect to the dense gas fraction derived from HCN is not constant from normal galaxies to LIRGs, ULIRGs and high-z gas rich galaxies.}
         \label{ulirgs-2}
   \end{figure}

However, as argued in the previous section, the use of HCN as an unbiased tracer of dense molecular gas in LIRGs and ULIRGs has been questioned on several fronts. First, X-rays may significantly enhance HCN abundances in enshrouded AGN, as suggested by the results obtained in nearby active galaxies. 
Furthermore the excitation of HCN lines in LIRGs and ULIRGs might be affected by IR pumping through a 14\,$\mu$m vibrational transition \citep{aal95, gb06}. In summary, the chemistry and the excitation of HCN lines in ULIRGs with embedded AGNs are suspected to depart from normal galaxy standards. Taking together, the possible caveats on the use of HCN observations call for the use of alternative tracers of dense gas in LIRGs and ULIRGs. This question is central to disentangling the different power sources of the huge infrared luminosities of these galaxies.

\begin{figure*}[!th]
  \centering
    \includegraphics[width=13.5cm]{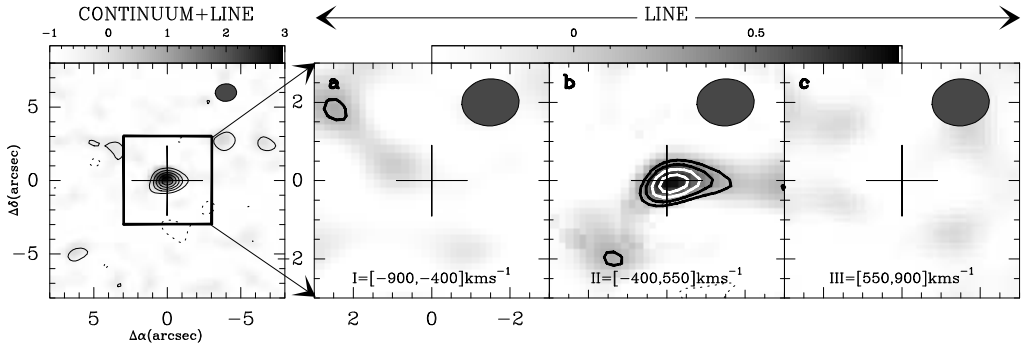}
     \caption{The first detection of HCO$^+$ emission at redshift z$\sim$4 in the QSO APM 08279+5255 \citep{gb06}. We show the continuum+line emission (left panel) and HCO$^+$(5--4) line emission in the three velocity intervals shown (right panels).}
         \label{apm}
   \end{figure*}

\citet{gra06} completed recently observations with the IRAM 30m telescope in the 1--0 line of HCO$^+$ of a sample of 16 galaxies including 10 LIRGs and 6 ULIRGs. Preliminary results of this HCO$^+$ survey, the first ever conducted in LIRGs and ULIRGs, indicate that the HCN/HCO$^+$ luminosity ratio increases with $L_{\rm{IR}}$ for LIRGs and ULIRGs (Fig.~\ref{ulirgs-1}). This unexpected trend provides indicative evidence that HCN is not a fair tracer of dense gas in the most extreme LIRGs.  In particular, the application to our sample of the diagnostic tool originally designed by \citet{koh05} to distinguish between `pure' AGNs and `composite' starbursts+AGNs in nearby Seyferts, suggests that a large number of embedded AGNs may lie in LIRGs and ULIRGs. A plausible scenario accounting for the observed trends implies that X-rays may shape the chemistry of molecular gas at $L_{\rm{IR}} > 10^{12}\,L_{\odot}$. Alternatively, it has also been argued that the abundance of HCN can be enhanced in the molecular gas closely associated with high-mass star forming regions.  In this case the reliability of HCN as a straightforward tracer of dense molecular gas in ULIRGs should be equally put on hold, even if the star formation scenario applies here \citep{gra06}.

 Different mechanisms, either related to the excitation of the HCN(1--0) line, or to the chemical enhancement of the HCN molecule 
can make for HCN(1--0) being over-luminous with respect to HCO$^+$(1--0). The caveats on the interpretation of HCN observations highlight the need of surveys in other molecular species that together provide an unbiased estimate of the dense molecular gas fraction of LIRGs and ULIRGs. The relevance of the understanding of ULIRGs resides in the fact that they may represent the local examples of the high redshift galaxies that dominate the IR and submm backgrounds. 

 The validity of \citet{gao04} 's paradigm in which SFE in galaxies (measured with respect to the dense gas content) is constant as function of $L_{\rm{IR}}$, has been recently examined by Graci\'a-Carpio, Garc\'{\i}a-Burillo \& Planesas, (2007; in prep), through the addition of new HCN(1-0) observations made in a sample LIRGs and ULIRGs. Figure~\ref{ulirgs-2} shows the first results of this new HCN(1--0) 30m-telescope survey. Although the scatter is still large after the addition of these new data, Figure~\ref{ulirgs-2} shows tantalizing evidence that the SFE measured with respect to the dense gas fraction is not constant as a function of $L_{\rm{IR}}$, but, on the contrary, it is significantly higher in some ULIRGs and high-z galaxies compared to the average value for normal galaxies. Alternatively, this result can be interpreted in terms of a significant AGN contribution to $L_{\rm{IR}}$ in some ULIRGs and high-z galaxies.  It is tempting to foresee that a similar SFE plot derived from HCO$^{+}$ data would even increase the differences between galaxies as a function of $L_{\rm{IR}}$, according to the findings of \citet{gra06}.

\subsection{Probing AGN feedback at z$\sim$4}

The use of bona fide tracers of the dense molecular gas phase in high redshift galaxies is of paramount importance for our understanding of the first phases of galaxy formation. While CO lines are commonly used to trace the general molecular gas 
phase at high-redshift, it is only recently that more complex molecular species have started to be used to study the dense molecular content in these galaxies.

The broad absorption line quasar APM~08279+5255 at z=3.911 is one of the most luminous sources in the universe even after correcting for the high lensing factor of its huge measured infrared luminosity (L$_{IR}\sim$10$^{15}$L$_{\odot}$). It is also a clear example of dominant AGN contribution 
to the total L$_{IR}$ budget.  Emission of high-J CO lines (J=9--8 and 4--3), mapped by \citet{dow99}, previously revealed the presence of a circumnuclear disk of hot and dense molecular gas. The question of how much dense molecular gas lies inside APM~08279+5255 was later addressed by \citet{wag05} who reported the detection of HCN~(5--4) emission in this quasar using the IRAM PdBI.  The exceptionally strong intensity of the HCN~(5--4) line with respect to all CO lines measured in APM~08279+5255 came out as a surprise. \citet{wag05} argue that the high HCN/CO luminosity ratios could stem from a large enhancement of the abundance of HCN relative to CO:  [HCN/CO]$\sim$(1--2)$\times$10$^{-2}$. This is an order of magnitude larger than the typical abundance ratio measured on small scales in galactic hot cores. Most notably, in the case of APM~08279+5255, the inferred [HCN/CO] ratio would correspond to a $\sim$10$^{10}$M$_{\odot}$ molecular gas disk. This is a good illustration of how extreme molecular gas chemistry can be at high redshift.

\citet{gb06} recently reported the detection of HCO$^+$~(5--4) emission from APM~08279+5255 based on observations conducted at the IRAM PdBI (see Fig.~5). This represents the first detection of this molecular ion at such a high redshift (see also detection of HCO$^{+}$(1-0) emission in the Cloverleaf by \citet{rie06}). The HCO$^+$ line profile central velocity and width are consistent with those derived from HCN. This result suggests that HCO$^+$~(5--4) emission comes roughly from the 
same circumnuclear region probed by HCN. However the HCN~(5--4)/HCO$^+$~(5--4) luminosity ratio measured by \citet{gb06} in APM~08279+5255$\sim$1.1 is $\sim$3 larger than that predicted by simple radiative transfer models which assume that excitation of the two lines is collisional and that chemical abundances for the two molecular species are comparable in this object.
In other words, if we stay inside the collisional excitation model, the abundance of HCN in APM~08279+5255 would be {\it anomalously} high with respect to CO, but also with respect to HCO$^+$: [HCN/CO]$\sim$(1--2)$\times$10$^{-2}$ and [HCN/HCO$^{+}$]$\sim$10.  As argued in the previous section, different mechanisms can make for HCN being over-abundant with respect to CO and HCO$^+$. This includes high-ionization chemistry driven by X-rays around an AGN and, alternatively, chemical enhancement of HCN in star-forming regions. Both ingredients (a massive star forming episode and an AGN) are at work in APM~08279+5255 and can probably drive the chemistry of a large amount of molecular gas.

Alternatively, \citet{gb06} argue that the excitation of HCN and HCO$^+$ lines in a source with an infrared luminosity as high as that of APM~08279+5255 may not be only collisional, but also radiative. Future observations that include lower and higher-J transitions of both HCO$^+$ and HCN will be required to confront a set of observed line ratios with the predictions issued from the two excitation schemes. Of particular note, the two scenarios invoked above, both accounting for the {\it unexpectedly} large
HCN(5--4)/HCO$^+$(5--4) ratio measured in APM~08279+5255, have completely different but equally relevant 
implications for the interpretation of high-J molecular line observations of dense gas in other high-redshift galaxies.

\label{}





\end{document}